\documentclass[aps,amsmath,amssymb,preprint]{revtex4}

\begin{document}

\title{A high-order electromagnetic gyrokinetic model}
\author{N. Miyato}
\email{miyato.naoaki@jaea.go.jp}
\affiliation{Japan Atomic Energy Agency, 2-116 Omotedate, Obuchi, Rokkasho,
Aomori, 039-3212 Japan}
\date{August 2013}%

\begin{abstract}
A high-order extension is presented for the electromagnetic gyrokinetic 
formulation in which the parallel canonical momentum is taken as one of 
phase space coordinates.
The high-order displacement vector associated with the guiding-center transformation
should be considered in the long wavelength regime.
This yields addtional terms in the gyrokinetic Hamiltonian which 
lead to modifications to the gyrokinetic Poisson and Amp\`ere equations.
In addition, the high-order piece of the guiding-center transformation
for the parallel canonical momentum should be also kept in the electromagnetic model.
The high-order piece contains the Ba\~nos drift effect and further modifies
the gyrokinetic Amp\`ere equation.
\end{abstract}

\maketitle



\section{Introduction}
The gyrokinetic models are widely used for studies of low frequency microturbulence in strongly 
magnetized plasmas.
The standard gyrokinetic model is formulated for perturbations
with small amplitude ($e\varphi/T\sim\epsilon_\delta\ll 1$) and short wavelength ($k_\perp\rho\sim 1$)
(gyrokinetic ordering)\cite{Hahm88,Brizard89,Brizard-Hahm07},
where $\varphi$ is the electrostatic potential,
$k_\perp$ the perpendicular wavenumber and $\rho$ the Larmor radius.
The gyrokinetic ordering can be interpreted as 
the slow flow condition 
$e{\boldsymbol\rho}\cdot\nabla_\perp\varphi/T\sim V_E/V_{\rm th}\ll 1$
with the {\bf E}$\times${\bf B} drift velocity $V_E$ and 
the thermal velocity $V_{\rm th}$.
Although this slow flow condition is also satisfied in the long wavelength regime
($k_\perp\rho\ll1, e\varphi/T\sim1$),
the standard gyrokinetic quasi-neutrality equation which is truncated to $O(\epsilon_\delta)$
is not always valid in the long wavelength regime.
For the long wavelength component of $\varphi$, the polarization term with $\varphi$
can go to the higher order and then the other higher order terms should be 
kept in the Poisson or quasi-neutrality equation.
The high-order displacement vector associated with the guiding-center transformation
gives the other higher order terms 
and should be kept in the long wavelength regime\cite{Miyato13}.
Since the guiding-center model is constructed 
up to high order enough by Littlejohn\cite{Littlejohn81,Littlejohn83}, 
it is no necessary to recalculate the guiding-center model.
Although large electric field is considered in the original Littlejohn guiding-center model,
related terms are neglected in the standard gyrokinetic model. 
Another guiding-center model with large electric field is found in
\cite{Miyato09}.
When the large electric field is neglected, the high-order displacement vector
is related to the nonuniformity of magnetic field only. 
Therefore, the high-order terms may be important for
the components whose wavelengths are 
comparable to that of the magnetic field.

The high-order contributions are not considered at the gyro-center
transformation stage in the standard gyrokinetic formulation
since they are negligible for the short wavelength perturbations.
The gyro-center models with the high-order contributions 
for general electromagnetic perturbations are
found in \cite{BrizardPHD} in which the parallel velocity $v_\parallel$ is 
an independent variable.
The calculation of the $v_\parallel$ formulation
with magnetic perturbations is rather cumbersome.
This is because the gyro-center transformation becomes 
complicated due to the vector potential perturbation at the symplectic 
part of the fundamental 1-form or the phase space Lagrangian.
There is no such complication in the electrostatic limit and
the high-order gyrokinetic model is constructed easily\cite{Miyato13}.
The simplication 
is achieved even in the electromagnetic case 
by taking the parallel canonical momentum $p_\parallel$ 
as an independent variable instead of $v_\parallel$
\cite{Hahm-Lee-Brizard88,Hahm09}.
If only the shear-Alfv\'enic fluctuations ($A_\parallel$) are considered,
the use of $p_\parallel$ deletes $A_\parallel$ from the symplectic part.
In this paper we extend the high-order gyrokinetic model in the electrostatic
limit to the electromagnetic one in terms of $p_\parallel$.

This paper is organized as follows.
The well-known guiding-center model with $p_\parallel$ is briefly explained 
in Sec. II.
In Sec. III, we derive the gyrokinetic Hamiltonian with additional terms
related to the high-order pieces in the guiding-center transformation.
The gyrokinetic Poisson and Amp\`ere equations are obtained 
systematically through the functional derivatives of 
the derived gyrokinetic Hamiltonian in Sec. IV.
Finally a summary is given in Sec. V.

\section{Guiding-center transformation}

The fundamental 1-form for a charged particle
with mass $m$ and electric charge $q$  
in an equilibrium magnetic field with small electromagnetic perturbations
($\phi$, $A_\parallel$)
is written in the particle phase space as 
\begin{eqnarray}
\gamma
=
\left[q{\bf A}_0({\bf x})+\epsilon_\delta qA_\parallel({\bf x},t)\hat{b}+m{\bf v}\right]
\cdot d{\bf x}
-\left[\epsilon_\delta q\phi({\bf x},t)+
\frac{m}{2}|{\bf v}|^2
\right]dt,
\label{eq:1-form}
\end{eqnarray}
where 
${\bf x}$ and ${\bf v}$ are the particle position and velocity, respectively, and 
$\hat{b}={\bf B}_0/|B_0|$ is the unit vector along the equilibrium magnetic field
 ${\bf B}_0\equiv\nabla\times{\bf A}_0$.
Only the shear Alfv\'en perturbation is considered here.
Changing the velocity variables to $(v_\perp,v_\parallel,\theta)$ gives
\begin{eqnarray}
\gamma
&=&
\left[
q{\bf A}_0({\bf x})+\epsilon_\delta qA_\parallel({\bf x},t)\hat{b}
+mv_\parallel\hat{b}+mv_\perp\hat{c}
\right]
\cdot d{\bf x} \nonumber \\
& &
-\left[\epsilon_\delta q\phi({\bf x},t)+
\frac{mv_\parallel^2}{2}+\frac{mv_\perp^2}{2B_0}B_0
\right]dt,
\end{eqnarray}
where 
$\hat{c}=-\sin\theta\hat{e}_1-\cos\theta\hat{e}_2$ 
is the unit vector along the velocity vector
perpendicular to $\hat{b}$, 
and $\hat{e}_1$ and $\hat{e}_2$ are unit vectors spanning the perpendicular plane.
The symplectic part of the above 1-form has the time dependency
through $A_\parallel$.
Therefore, the phase space transformation to remove fast gyromotion 
from the 1-form becomes more complicated compared to the electrostatic 
case.
There is an easy solution for this.
When, as one of phase space coordinates, we use the parallel canonical momentum,
\begin{equation}
p_\parallel\equiv mv_\parallel+qA_\parallel,
\end{equation}
$A_\parallel$ disappears from the symplectic part as,
\begin{eqnarray}
\gamma
&=&
\left[
q{\bf A}_0({\bf x})
+p_\parallel\hat{b}+mv_\perp\hat{c}
\right]
\cdot d{\bf x} \nonumber \\
& &
-\left[\epsilon_\delta q\phi({\bf x},t)+
\frac{(p_\parallel-\epsilon_\delta qA_\parallel)^2}{2m}+\frac{mv_\perp^2}{2B_0}B_0
\right]dt.
\end{eqnarray}
The time dependent perturbations appear only in 
the Hamiltonian.
The $O(\epsilon_\delta^0)$ part given by
\begin{eqnarray}
\gamma_0
=
\left[
q{\bf A}_0({\bf x})
+p_\parallel\hat{b}+mv_\perp\hat{c}
\right]
\cdot d{\bf x} 
-\left[\frac{p_\parallel^2}{2m}+\frac{mv_\perp^2}{2B_0}B_0\right]dt
\end{eqnarray}
is the usual unperturbed 1-form except $mv_\parallel\to p_\parallel$.
Therefore, the standard guiding-center transformation can be applied 
in order to remove the gyrophase dependence from the 1-form
and gives the following guiding-center 1-form\cite{Brizard-Hahm07},
\begin{equation}
\Gamma=
q{\bf A}^*\cdot d{\bf X}
+\frac{m}{q}\mu d\xi-\left[\frac{P_\parallel^2}{2m}+\mu B_0\right]dt,
\end{equation}
where ${\bf Z}=({\bf X}, P_\parallel, \mu, \xi)$ are the guiding-center coordinates,
${\bf A}^*$ is the modified vector potential given by
\begin{equation}
{\bf A}^*={\bf A}_0+\frac{P_\parallel}{q}\hat{b}-\frac{m}{q^2}\mu{\bf W},
\end{equation}
and ${\bf W}=(\nabla\hat{e}_1)\cdot\hat{e}_2+(\hat{b}\cdot\nabla\times\hat{b})\hat{b}/2$.

\section{Gyrokinetic Hamiltonian}

Now we consider the perturbations which have still the gyrophase dependence.
In order to remove the remaining gyrophase dependence, 
the gyro-center transformation will be performed.
As shown in the previous section, the perturbations only appear in the Hamiltonian.
Hence, no modification to the symplectic part is needed.
In this case the gyro-center transformation becomes the simple
canonical transformation and
only the Hamiltonian is modified.
The perturbed Hamiltonian in terms of the guiding-center coordinates is
formally represented by
\begin{equation}
h\left({\bf Z},t\right)
=
q\left[
\phi({\sf T}^{-1}_{\rm GC}{\bf x}, t)
-\frac{{\sf T}^{-1}_{\rm GC}p_\parallel}{m}A_\parallel({\sf T}^{-1}_{\rm GC}{\bf x}, t)
\right]
+
\frac{q^2}{2m}A_\parallel^2({\sf T}^{-1}_{\rm GC}{\bf x}, t),
\end{equation}
where ${\sf T}^{-1}_{\rm GC}{\bf x}$ and ${\sf T}^{-1}_{\rm GC}p_\parallel$
denote the particle position and particle parallel momentum in the guiding-center
phase space, respectively.
In the standard formulation ${\sf T}^{-1}_{\rm GC}{\bf x}$ is approximated by
${\sf T}^{-1}_{\rm GC}{\bf x}={\bf X}+{\boldsymbol\rho}$.
When we consider the long wavelength regime, however,
the higher-order displacement vector
should be retained in ${\sf T}^{-1}_{\rm GC}{\bf x}$
as
\begin{equation}
{\sf T}^{-1}_{\rm GC}{\bf x}
={\bf X}+{\boldsymbol\rho}+{\boldsymbol\rho}_B.
\end{equation}
We denotes the high-order displacement vector by
${\boldsymbol\rho}_B$ which is in general defined by
\begin{equation}
{\boldsymbol\rho}_B
\equiv
-\left(G_2^{\bf X}-\frac{1}{2}{\bf G}_1\cdot{\bf d}G_1^{\bf X}\right),
\label{eq:rho_B}
\end{equation}
where ${\bf G}_n$ is the vector field generating the guiding-center
transformation at $n$th order, 
and ${\bf G}_n\cdot{\bf d}=G_n^j\partial_j$.
The usual gyroradius vector is given by
${\boldsymbol\rho}=-G_1^{\bf X}$. 
The explicit representation of ${\boldsymbol\rho}_B$ is found 
in \cite{Littlejohn81,Miyato13,Miyato11}.
Considering ${\boldsymbol\rho}_B$ in ${\sf T}^{-1}_{\rm GC}{\bf x}$,
we may expand the potentials as 
\begin{equation}
\phi({\sf T}^{-1}_{\rm GC}{\bf x})
\simeq\phi({\bf X}+{\boldsymbol\rho})
+{\boldsymbol\rho}_B\cdot\nabla\phi({\bf X}+{\boldsymbol\rho}),
\end{equation}
\begin{equation}
A_\parallel({\sf T}^{-1}_{\rm GC}{\bf x})
\simeq A_\parallel({\bf X}+{\boldsymbol\rho})
+{\boldsymbol\rho}_B\cdot\nabla A_\parallel({\bf X}+{\boldsymbol\rho}).
\end{equation}
Similarly, although in the standard model ${\sf T}^{-1}_{\rm GC}p_\parallel$ is simply
 replaced by the lowest order term $P_\parallel$,
we retain here the higher order term for ${\sf T}^{-1}_{\rm GC}p_\parallel$
\begin{equation}
{\sf T}^{-1}_{\rm GC}p_\parallel=P_\parallel-G_1^{P_\parallel},
\end{equation}
where $G_1^{P_\parallel}$ is the $p_\parallel$ component of ${\bf G}_1$ given by
\begin{equation}
G_1^{P_\parallel}=\frac{m\mu}{q}({\sf a}_1:\nabla\hat{b}+\hat{b}\cdot\nabla\times\hat{b})
-P_\parallel{\boldsymbol\rho}\cdot(\hat{b}\cdot\nabla\hat{b}),
\end{equation}
with ${\sf a}_1=-(\hat{a}\hat{c}+\hat{c}\hat{a})/2$, 
$\hat{a}=\hat{b}\times\hat{c}$.
Then, the peturbed Hamiltonian in the guiding-center phase space
is written as
\begin{eqnarray}
h\left({\bf Z},t\right)
&=&
q\left[
\phi({\sf T}^{-1}_{\rm GC}{\bf x})-\frac{P_\parallel-G_1^{P_\parallel} }{m}A_\parallel
({\sf T}^{-1}_{\rm GC}{\bf x})
\right]
+
\frac{q^2}{2m}A_\parallel^2({\sf T}^{-1}_{\rm GC}{\bf x})
\nonumber \\
&=&
q\left[
\phi({\sf T}^{-1}_{\rm GC}{\bf x})-\frac{P_\parallel}{m}A_\parallel
({\sf T}^{-1}_{\rm GC}{\bf x})
\right]
+
\frac{q}{m}G_1^{P_\parallel}A_\parallel
+
\frac{q^2}{2m}A_\parallel^2({\sf T}^{-1}_{\rm GC}{\bf x}).
\end{eqnarray}
The phase space transformation to the gyro-center coordinates
$\bar{\bf Z}=(\bar{\bf X},\bar{P}_\parallel,\bar{\mu},\bar{\xi})$
is performed to remove gyrophse dependence
from the above perturbed Hamiltonian.
The lowest order gyro-center Hamiltonian is simply 
$\bar{H}_0=\bar{P}_\parallel^2/2m+\bar{\mu}B_0$.
The perturbed Hamiltonian is
given by
\begin{eqnarray}
\bar{h}
&=&
q(\langle\psi(\bar{\bf X}+\bar{\boldsymbol \rho})\rangle
+\langle\bar{\boldsymbol\rho}_B\cdot\bar{\nabla}
\psi(\bar{\bf X}+\bar{\boldsymbol \rho})\rangle)
\nonumber \\
& &
+\frac{q}{m}\langle G_1^{P_\parallel}(\bar{\bf Z})
A_\parallel(\bar{\bf X}+\bar{\boldsymbol \rho})\rangle
+\frac{q^2}{2m}
\langle A_\parallel^2(\bar{\bf X}+\bar{\boldsymbol \rho})\rangle
-\frac{q}{2}\langle\{S_1,\tilde{\psi}\}\rangle,
\end{eqnarray}
where $\langle\cdot\rangle$ denotes the gyrophase average 
and $\psi$ is the generalized potential defined by
\begin{equation}
\psi(\bar{\bf Z},t)\equiv\phi(\bar{\bf Z},t)
-\frac{\bar{P}_\parallel}{m}A_\parallel(\bar{\bf Z},t).
\end{equation}
The scalar function generating the first
order gyro-center transformation is
\begin{equation}
S_1=\frac{q}{\Omega}\int\tilde{\psi}d\bar{\xi},
\end{equation}
where 
$\tilde{\psi}=\psi(\bar{\bf X}+\bar{\boldsymbol\rho})-\langle\psi(\bar{\bf X}+\bar{\boldsymbol\rho})\rangle$ 
is the oscillatory part of $\psi$.
The nonlinear term of $\psi$ is usually approximated by
\begin{equation}
\langle\{S_1,\tilde{\psi}\}\rangle
\simeq
\frac{q^2}{m\Omega}\frac{\partial\langle\tilde{\psi}^2\rangle}{\partial\bar{\mu}}
=\frac{q^2}{m\Omega}\frac{\partial}{\partial\bar{\mu}}
(\langle\psi^2\rangle-\langle\psi\rangle^2).
\end{equation}
The terms with ${\boldsymbol\rho}_B$ and $G_1^{P_\parallel}$
are only important in the long wavelength regime.
Hence, the perturbed Hamiltonian may be approximated as
\begin{eqnarray}
\bar{h}(\bar{\bf X},\bar{P}_\parallel,\bar{\mu},t)
&=&
q(\langle\psi(\bar{\bf X}+\bar{\boldsymbol \rho})\rangle
+\langle\bar{\boldsymbol\rho}_B\rangle\cdot\bar{\nabla}
\psi(\bar{\bf X})+\frac{q}{m}\langle G_1^{P_\parallel}(\bar{\bf Z})\rangle
A_\parallel(\bar{\bf X})
\nonumber \\
& &
+\frac{q^2}{2m}
\langle A_\parallel^2(\bar{\bf X}+\bar{\boldsymbol \rho})\rangle
-\frac{q^2}{2B_0}\frac{\partial}{\partial\bar{\mu}}
(\langle\psi^2\rangle-\langle\psi\rangle^2).
\label{eq:H_gk}
\end{eqnarray}
where $\langle{\boldsymbol\rho}_B\rangle$ and $\langle G_1^{P_\parallel}\rangle$
are, respectively, given by 
\begin{eqnarray}
\langle{\boldsymbol\rho}_B\rangle
=
-\left\{
\frac{\mu B_0}{m\Omega^2}\frac{1}{2}(\nabla\cdot\hat{b})\hat{b}
+\frac{U^2}{\Omega^2}\hat{b}\cdot\nabla\hat{b}
+\frac{3}{2}\frac{\mu B_0}{m\Omega^2}\nabla_\perp\log B_0
\right\},
\label{eq:rho_B_average}
\end{eqnarray}
and
\begin{equation}
\langle G_1^{P_\parallel}\rangle
=\frac{m\mu}{q}\hat{b}\cdot\nabla\times\hat{b}.
\end{equation}

\section{Field equations}

The gyrokinetic field equations are easily obtained 
through the field-theoretical treatment\cite{Sugama00}.
The Poisson equation is given by
\begin{equation}
\epsilon_0\nabla^2\phi({\bf r})=-\sum_{\rm sp}\int d^6\bar{\bf Z}\bar{\cal J}\bar{F}
\frac{\delta\bar{h}(\bar{\bf Z})}{\delta\varphi({\bf r})},
\label{eq:Poisson_general}
\end{equation}
where $\epsilon_0$ is the permittivity of vacuum,
$\bar{\cal J}=B_\parallel^*/m^2$, $B_\parallel^*\equiv\hat{b}\cdot{\bf B}^*$,
${\bf B}^*\equiv\nabla\times{\bf A}^*$, 
and the summation is taken over species.
Similarly, the gyrokinetic Amp\`ere equation is given by
\begin{equation}
\frac{1}{\mu_0}\nabla_\perp^2A_\parallel({\bf r})
=\sum_{\rm sp}\int d^6\bar{\bf Z}\bar{\cal J}\bar{F}
\frac{\delta\bar{h}(\bar{\bf Z})}{\delta A_\parallel({\bf r})},
\label{eq:Ampere_general}
\end{equation}
where $\mu_0$ is the permeability of vacuum.
It is noted that the lowest order Hamiltonian does not have the potential perturbations
and therefore only the functional derivatives of the perturbed Hamiltonian
appear in the field equations.
Taking the functional derivatives of the perturbed Hamiltonian (\ref{eq:H_gk}),
we have
\begin{eqnarray}
\frac{\delta\bar{h}(\bar{\bf Z})}{\delta\varphi({\bf r})}
&=&
q\langle\delta^3(\bar{\bf X}+\bar{\boldsymbol\rho}-{\bf r})\rangle
+q\langle\bar{\boldsymbol\rho}_B\rangle\cdot\bar{\nabla}\delta^3(\bar{\bf X}-{\bf r})
\nonumber \\
& &
-\frac{q^2}{B_0}\frac{\partial}{\partial\bar{\mu}}
(\langle\psi(\bar{\bf X}+\bar{\boldsymbol\rho})\delta^3(\bar{\bf X}+\bar{\boldsymbol\rho}-{\bf r})\rangle
-\langle\psi(\bar{\bf X}+\bar{\boldsymbol\rho})\rangle
\langle\delta^3(\bar{\bf X}+\bar{\boldsymbol\rho}-{\bf r})\rangle)
\label{eq:delHdelphi}
\end{eqnarray}
and
\begin{eqnarray}
\frac{\delta\bar{h}(\bar{\bf Z})}{\delta A_\parallel({\bf r})}
&=&
-\frac{q\bar{P}_\parallel}{m}\langle\delta^3(\bar{\bf X}+\bar{\boldsymbol\rho}-{\bf r})\rangle
-\frac{q\bar{P}_\parallel}{m}\langle\bar{\boldsymbol\rho}_B\rangle\cdot\bar{\nabla}\delta^3(\bar{\bf X}-{\bf r})
+\frac{q}{m}\langle G_1^{P_\parallel}\rangle\delta^3(\bar{\bf X}-{\bf r})
\nonumber \\
& &
+\frac{q^2}{m}\langle A_\parallel(\bar{\bf X}+\bar{\boldsymbol\rho})\delta^3(\bar{\bf X}+\bar{\boldsymbol\rho}-{\bf r})\rangle
\nonumber \\
& &
+\frac{q^2\bar{P}_\parallel}{mB_0}\frac{\partial}{\partial\bar{\mu}}
(\langle\psi(\bar{\bf X}+\bar{\boldsymbol\rho})\delta^3(\bar{\bf X}+\bar{\boldsymbol\rho}-{\bf r})\rangle
-\langle\psi(\bar{\bf X}+\bar{\boldsymbol\rho})\rangle
\langle\delta^3(\bar{\bf X}+\bar{\boldsymbol\rho}-{\bf r})\rangle).
\label{eq:delHdelA}
\end{eqnarray}
Substituting Eq. (\ref{eq:delHdelphi}) into Eq. (\ref{eq:Poisson_general}) 
and integrating by parts, we have the gyrokinetic Poisson equation
\begin{eqnarray}
\epsilon_0\nabla^2\varphi({\bf r})
&=&
-\sum_{\rm sp}q\Bigg[
\int d^6\bar{\bf Z}
\left(
\bar{F}\bar{\cal J}
+\frac{q\tilde{\psi}}{B_0}\frac{\partial\bar{F}\bar{\cal J}}{\partial\bar{\mu}}
\right)
\delta^3(\bar{\bf X}+\bar{\boldsymbol\rho}-{\bf r})
\nonumber \\
& &
-\int d^6\bar{\bf Z}\delta^3(\bar{\bf X}-{\bf r})
\bar{\nabla}\cdot\bar{F}\bar{\cal J}\langle\bar{\boldsymbol\rho}_B\rangle
\Bigg].
\end{eqnarray}
The Poisson equation is the same as that in the electrostatic case
except that the potential on the right hand side is not $\varphi$ but $\psi$\cite{Miyato13}.
The last term on the right hand side is the additional term due to ${\boldsymbol\rho}_B$.
Similarly, substituting Eq. (\ref{eq:delHdelA}) into Eq. (\ref{eq:Ampere_general}) 
and integrating by parts, we have the gyrokinetic Amp\`ere equation
\begin{eqnarray}
\frac{1}{\mu_0}\nabla_\perp^2A_\parallel({\bf r})
&=&
-\sum_{\rm sp}q\Bigg[\int d^6\bar{\bf Z}
\left(
\frac{\bar{P}_\parallel-qA_\parallel}{m}
\bar{F}\bar{\cal J}
+\frac{\bar{P}_\parallel q\tilde{\psi}}{mB_0}
\frac{\partial\bar{F}\bar{\cal J}}{\partial\bar{\mu}}
\right)
\delta^3(\bar{\bf X}+\bar{\boldsymbol\rho}-{\bf r})
\nonumber \\
& &
-\int d^6\bar{\bf Z}\delta^3(\bar{\bf X}-{\bf r})
\left(
\frac{\bar{P}_\parallel}{m}\bar{\nabla}\cdot\bar{F}\bar{\cal J}
\langle\bar{\boldsymbol\rho}_B\rangle
+\bar{F}\bar{\cal J}\frac{\langle G_1^{P_\parallel}(\bar{\bf Z})\rangle}{m}
\right)
\Bigg].
\label{eq:Ampere}
\end{eqnarray}
It is seen that besides the term with ${\boldsymbol\rho}_B$,
the term due to 
$\langle G_1^{P_\parallel}(\bar{\bf Z})\rangle=(m\bar{\mu}/q)\hat{b}\cdot\bar{\nabla}\times\hat{b}$
appears in the Amp\`ere equation.
This shows the effect of the Ba\~nos term on the parallel current density\cite{Banos67}.
The explicit form of this term may change by another choice for $G_1^{P_\parallel}$
\cite{Littlejohn82,Brizard89}.
The appearance of this term is also shown by considering 
the push-forward representation of the parallel current density 
associated with the guiding-center transformation.
The right hand side of the Amp\`ere equation (\ref{eq:Ampere})
must be the parallel {\it particle} current density.
Hence, it can be regarded as the push-forward representation of the 
parallel current density which
is written rigorously as
\begin{eqnarray}
j_\parallel({\bf r})
&=&
\sum_{\rm sp}q\int d^3{\bf x}d^3{\bf v}
v_\parallel f({\bf x},{\bf v})\delta^3({\bf x}-{\bf r})
\nonumber \\
&=&
\sum_{\rm sp}q\int d^6{\bf Z}{\cal J}
{\sf T}_{\rm GC}^{-1}v_\parallel F({\bf Z})\delta^3({\sf T}_{\rm GC}^{-1}{\bf x}-{\bf r}),
\end{eqnarray}
where 
${\sf T}_{\rm GC}^{-1}v_\parallel=U-G_1^U+\cdots$ denotes 
the particle parallel velocity in the guiding-center phase space
and $U$ is the guiding-center parallel velocity.
The gyroaverage of $\langle G_1^U\rangle$ agrees with
$\langle G_1^{P_\parallel}\rangle/m$.
The additional term due to ${\boldsymbol\rho}_B$ stems from 
${\sf T}_{\rm GC}^{-1}{\bf x}$ in the delta function.

\section{Summary}

We reformulated the electromagnetic gyrokinetic model with
the high-order pieces associated with the guiding-center transformation
which are not considered in the standard gyrokinetic formulations.
The use of parallel canonical momentum $p_\parallel$ instead of 
parallel velocity $v_\parallel$ deletes the magnetic perturbation $A_\parallel$
from the symplectic part of 1-form, and thereby
makes it easier to extend the high-order gyrokinetic model 
to the electromagnetic one.
Not only the high-order displacement vector
${\boldsymbol\rho}_B$ but also $G_1^{P_\parallel}$ should be retained
in the electromagnetic case.
We derived the gyrokinetic Hamiltonian including the high-order pieces.
The field equations were easily obtained from the derived
gyrokinetic Hamiltonian through the variational method.
The additional term due to ${\boldsymbol\rho}_B$ in the gyrokinetic 
Poisson equation is the same as the one in the electrostaic limit.
$G_1^{P_\parallel}$ as well as ${\boldsymbol\rho}_B$ yields the additional term
in the gyrokinetic Amp\`ere equation, which contains the Ba\~nos drift effect
on the parallel current density.
The appearance of such term is also confirmed by simple consideration of the
push-forward representation of parallel current density.

\begin{acknowledgements}
This work is partly supported by KAKENHI (22760663)
from Japan Society for the Promotion of Science.
\end{acknowledgements}

\end{document}